# Molecular Tuning of Charge-Transfer Resonance in Plasmonic Nanocavities

Hasher Irshad[a,b], Francesco Ciccarello[b], Konstantin Malchow[b], Christophe Galland[b,*], Katrine Qvortrup[a,*]

[a] *Department of Chemistry, Technical University of Denmark, 2800, Kongens Lyngby, Denmark*

[b] *Institute of Physics, Ecole Polytechnique Fédérale de Lausanne (EPFL), CH-1015 Lausanne, Switzerland*

* *Corresponding authors*

**E-mail addresses:** chris.galland@epfl.ch (Christophe Galland) kaqvo@kemi.dtu.dk (Katrine Qvortrup)

## Abstract

Interfacial charge-transfer processes play a critical role in plasmon-enhanced spectroscopy, yet the energetic conditions governing charge-transfer resonance within molecule–metal nanocavities remain poorly understood. Here, plasmonic nanoparticle-on-mirror junctions incorporating systematically engineered biphenylthiol derivatives monolayers were used to investigate how frontier orbital alignment influences chemical enhancement mechanisms. A molecular library spanning a broad range of electronically tuned acceptor states was examined using surface-enhanced Raman scattering (SERS), vibrational sum frequency generation (vSFG) spectroscopy, and density functional theory calculations. By combining different excitation wavelengths with controlled variation of substrate composition and molecular electronic structure, the energetic relationship between plasmon-enhanced charge transfer excitation and molecular orbital alignment was quantitatively evaluated. The results reveal that charge-transfer enhancement of Raman scattering is governed by a well-defined interfacial resonance condition dependent on substrate work function and excitation energy. We further probe a subset of molecular-metal systems by nanocavity-enhanced vSFG and identify the same resonance conditions as in SERS, consistent with expectations. These findings establish an experimentally accessible framework for probing and engineering charge-transfer processes in plasmonic molecular junctions and provide mechanistic insight relevant to molecular plasmonics, charge-carrier photophysics, and nanoscale interfacial spectroscopy.

**Keywords**:

Charge-transfer resonance; Plasmonic nanocavities; Nanoparticle-on-mirror; Molecular orbital engineering; Chemical enhancement; SERS; vSFG

**Highlights:**

LUMO alignment with the substrate Fermi level and the excitation energy governs plasmonic charge-transfer resonance and chemical enhancement.

SERS and SFG intensity can be enhanced by CT.

Orbital engineering enables predictive enhancement.

**Graphical Abstract**

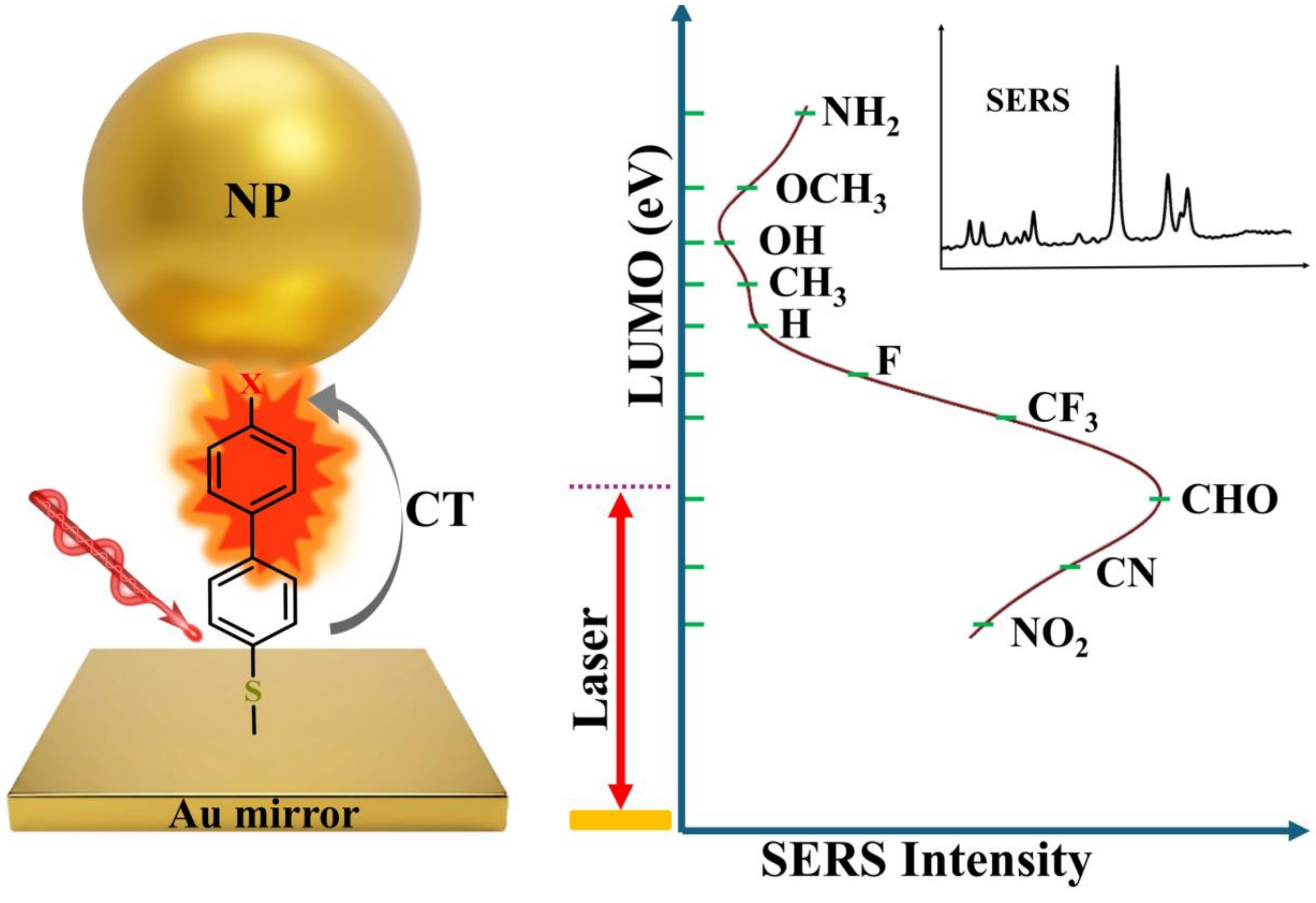

## 1. Introduction

Plasmonic nanocavities provide powerful nanomaterials for concentrating optical fields into deeply subwavelength volumes, enabling strong light–matter interactions at molecule–metal interfaces[1]. Among these architectures, nanoparticle-on-mirror (NPoM) systems have emerged as particularly attractive platforms because of their reproducible and confined nanogap geometry that simultaneously generates extreme electromagnetic field enhancement and direct electronic coupling between metallic surfaces and molecular junctions[2]. These properties have enabled their use in plasmon-enhanced spectroscopy[3], nanoscale catalysis[4], molecular optoelectronics[5], and plasmon-driven charge-transfer processes[6]. A common and desired feature of plasmon-molecular systems is the enhanced light-matter interaction, which makes it possible to perform incoherent[7,8] and coherent[9,10] vibrational spectroscopy on only one or a few molecules. Surface-enhanced Raman scattering (SERS) enhancement is commonly described as a combination of electromagnetic and chemical contributions[11].

While the plasmonic resonance being responsible for the electromagnetic enhancement in the near-field is well accepted, the origin of the chemical enhancement factor within the plasmonic cavities[12,13] remains not fully understood. The efficiency of chemical enhancement in plasmonic resonance depends sensitively on several critical nanoscale parameters, including molecule–metal electronic coupling, frontier orbital alignment, and excitation-energy-dependent interfacial transitions[14]. Chemical enhancement in SERS has been attributed to ground-state molecule–surface charge transfer[15], the Herzberg–Teller vibronic coupling[16], adsorption-induced changes in molecular polarizability[17,18], and binding-induced hybridization of molecular and substrate electronic states[19]. In particular, the interfacial charge transfer (CT) resonances between metallic substrates and molecular orbitals have long been suggested to contribute strongly to chemical enhancement[20]. However, experimental evidence on the energetic resonance conditions governing these processes has proven elusive because variations in molecular structure often simultaneously alter adsorption geometry, near-field confinement, packing density, vibrational selection rules, and intrinsic Raman activity[21]. Previous studies investigating substituent-dependent SERS enhancement have mostly relied on theoretical simulations, qualitative donor–acceptor descriptions, or limited molecular series, making it challenging to isolate the role of orbital energetics from competing structural variables[22]. Furthermore, although theoretical models predict that CT enhancement should be maximized when the excitation energy bridges the gap between the metal Fermi level and molecular acceptor states, direct experimental mapping of this resonance condition within highly confined plasmonic cavities remains limited[20,23,24]. The situation becomes even more complex in NPoM architectures, where strong confinement, image-charge effects, plasmon-induced charge-carrier distributions, and interfacial hybridization can substantially renormalize the effective electronic structure relative to idealized vacuum-level approximations[25,26].

To address these challenges, we developed a molecular library of systematically modified para-substituted biphenylthiol (BPT) derivatives assembled within plasmonic NPoM junctions. The molecular series was designed to have a consistent conjugated backbone and identical Au–thiolate anchoring geometry while continuously modifying the lowest unoccupied molecular orbital (LUMO) energy by installation of relevant substituents[27,28]. This strategy enables interrogation of how frontier orbital alignment influences plasmon-assisted interfacial charge transfer with minimal disturbance from structural heterogeneity within the nanocavity. Using a combination of SERS, dark-field scattering spectroscopy, vSFG spectroscopy, and density functional theory (DFT) calculations, we experimentally map the energetic dependence of CT-assisted enhancement in Au and Ag plasmonic nanocavities under 632 and 785 nm excitation[29]. To confront the structure-activity understanding gained from these studies, we rationally designed molecular derivatives possessing targeted frontier orbital energies and validated our predictions on these new molecules. Finally, interface-selective vSFG measurements were performed, which confirmed that the CT resonance condition identified with SERS at visible wavelengths is also addressable through coherent mid-infrared + visible vibrational spectroscopy. Together, these results provide an experimentally accessible framework for quantitatively mapping interfacial CT resonance conditions in plasmonic molecular junctions. More broadly, the work demonstrates that chemical enhancement within plasmonic nanocavities can be rationally engineered through precise control of molecular orbital alignment, providing mechanistic insight to plasmon-driven spectroscopy[30], charge-carrier photophysics[31], and nanoscale interfacial energy transfer[32].

## 2. Results and Discussion

### 2.1 System Design and Vibrational Signatures of Monolayers

To systematically investigate interfacial charge-transfer (CT) processes within strongly confined plasmonic nanocavities, nanoparticle-on-mirror (NPoM) junctions were constructed using self-assembled molecular monolayers on flat, template-stripped polycrystalline Au(111) substrates[33]. In this architecture, the target molecules are spatially confined within sub-nanometer plasmonic gaps formed between the metallic mirror substrate and the coupled nanoparticles, producing highly localized electromagnetic fields capable of amplifying both intra-molecular and interfacial optical transitions[34] (Fig. 1a). Such geometries provide an strong experimental platform for probing molecule–metal charge-transfer because the confined optical fields, well-defined nanogap environment, and direct metal–molecule contact collectively maximize the probability of plasmon-mediated electronic couplings[35].

A molecular library consisting of ten para-substituted biphenylthiol (BPT) derivatives was designed and synthesized to establish a systematically changed electronic framework for evaluating CT-mediated enhancement in NPoM junctions. The molecular series spans a broad range of biphenylthiols functionalized with electron-withdrawing and electron-donating substituents ($–NO_2$, –CN, –CHO, $–CF_3$, –F, –H, $–CH_3$, –OH, $–OCH_3$, and $–NH_2$), providing a library of structures with different lowest unoccupied molecular orbital (LUMO) energies while preserving the same conjugated biphenyl backbone and thiolate anchoring motif. This molecular design minimizes structural variables that commonly complicate comparative SERS studies, including differences in adsorption geometry, molecular footprint, plasmonic gap size, intermolecular packing, and electronic coupling to the metal surface[36,37]. Consequently, variations in optical responses can be interpreted predominantly in terms of substituent-induced electronic perturbations rather than geometric heterogeneity within the plasmonic cavity.

Before evaluating substituent-dependent intensity variations, we analyse normalized SERS spectra measured for each of the molecules in the molecular library. The compounds exhibited highly consistent vibrational fingerprints (Fig. 1b). The spectra are characterized by the symmetric aromatic ring breathing mode at 1005 $cm^{-1}$, the in-plane C–S stretching mode near 1080 $cm^{-1}$ associated with Au–thiolate bonding, the aromatic C–H bending mode at 1180 $cm^{-1}$, the inter-ring C–C stretching vibration at 1280 $cm^{-1}$, and the intense aromatic C=C stretching mode centred near 1585 $cm^{-1}$ [38]. The persistence and relative Raman intensities of these characteristic modes across all derivatives confirm the formation of structurally analogous molecular monolayers within the nanogap.

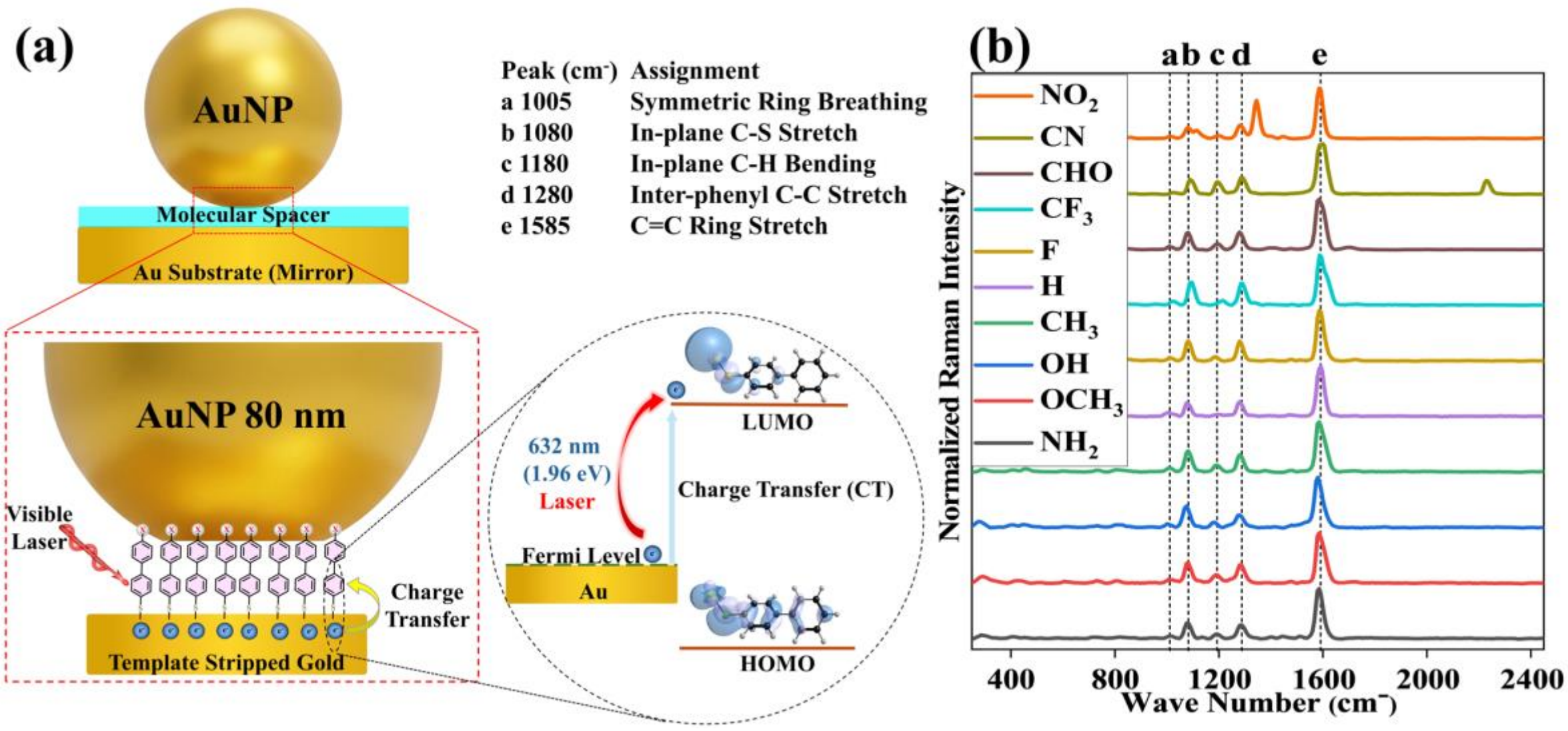

**Fig. 1 (a)** Schematic representation of the Au NPoM architecture and suggested charge transfer phenomenon. **(b)** Normalized SERS spectra of all ten BPT derivatives assembled in Au NPoM junctions under 632 nm excitation, highlighting the distinct vibrational responses.

The spectral uniformity observed across the BPT series and the absence of substantial peak shifts, spectral broadening, or mode suppression indicate that the molecular orientation, binding configuration, and overall junction geometry remain largely conserved despite significant electronic differences of the terminal substituents. This distinction is essential because it isolates electronic structure as the dominant experimental variable governing the observed Raman scattering enhancement. Therefore, differences in absolute Raman intensity can be primarily attributed to substituent-dependent modulation of interfacial electronic coupling and CT resonance conditions rather than alterations in the local electromagnetic field due to variations in molecular assembly or plasmonic cavity morphology[39].

Furthermore, the preservation of nearly identical vibrational signatures enables direct interrogation of how frontier orbital alignment influences chemical enhancement within the nanocavity. This controlled energetic tuning establishes the basis for quantitatively examining the relationship between molecular LUMO alignment and CT-mediated SERS intensity enhancement in subsequent sections. Additionally, density functional theory (DFT) calculations reproduced the experimental Raman vibrational fingerprints on Au-bound molecular models, confirming the structural consistency of the molecular series[40] (Fig. S1). The calculations provided frontier molecular orbital energies across the derivatives while having similar orbital topology and localization at the metal–molecule interface. The calculated LUMO distributions for all derivatives exhibited substantial orbital density localized near the Au–S anchoring region, suggesting that the unoccupied states remain electronically accessible to the metallic substrate and are therefore expected to have plausible acceptor states for substrate-to-molecule charge transfer under plasmonic excitation[41] (Fig. S2). Under optical excitation, plasmon-enhanced CT electronic transitions can be efficiently excited when the excitation energy approaches resonance with the effective energy separation between the metal Fermi level and the molecular LUMO[42]. Within this framework, modulation of substituent-dependent LUMO energies provides a direct strategy for experimentally mapping the energetic conditions that maximize CT-assisted enhancement[43].

**2.2 Quantitative CT Resonance Mapping and Laser Energy Dependence**

Having confirmed the structural consistency of the biphenylthiol monolayers within the plasmonic junctions, we examined the influence of the systematically modulated molecular LUMO energies on the SERS response of the Au NPoM cavities. Upon 632 nm excitation (1.96 eV), pronounced substituent-dependent intensity variations emerge across the molecular library while preserving nearly identical vibrational fingerprints and spectral positions (Fig. 1b and Fig. 2a). The Raman intensities of the characteristic biphenyl vibrational modes were plotted as a function of the calculated LUMO energies (Fig. 2a), revealing a clear non-monotonic dependency, with Raman enhancement increasing significantly for molecules possessing LUMO energies below approximately -3.4 eV and reaching a maximum for the –CHO substituted biphenyl having a calculated LUMO energy of -3.50 eV. Relative to the unsubstituted biphenyl (LUMO -3.31 eV), the CHO-functionalized biphenyl shows a 3.5-fold enhancement of the 1585 $cm^{-1}$ aromatic C=C stretching mode, a 4.3-fold enhancement of the 1280 $cm^{-1}$ inter-phenyl C–C stretching mode, and an approximately 4.1-fold enhancement for the 1080 $cm^{-1}$ Au–thiolate-associated C–S stretching vibration. The uniform SERS intensity enhancement observed across diverse vibrational modes indicates that any underlying chemical contribution is not localized, resulting in a homogeneous chemical enhancement across the entire molecular frame. Importantly, the observed enhancement trend does not align monotonically with conventional electron-withdrawing strength or substituent polarity. For example, although the –$NO_2$ derivative possesses the lowest LUMO energy in the series, it does not produce the highest Raman enhancement. Instead, the intensity reaches an optimum at intermediate energetic alignment near the –CHO derivative before decreasing again for molecules possessing progressively deeper LUMO levels.

Analysis of the broad SERS background also revealed a substituent-dependent response, with stronger backgrounds generally observed for electron-withdrawing derivatives (Fig. 2a and Fig S3). However, the values do not directly mirror the electron-withdrawing nature of the substituent, indicating that factors beyond the electronic factors must be influencing the mode-specific Raman enhancement. This supports the presence of molecule-dependent electronic coupling within the junction and is consistent with contributions from charge-transfer-enhanced SERS background[44]. Such broad continua in SERS have been attributed to plasmon-mediated emission and molecule–metal electronic coupling, rather than to a purely arbitrary baseline[45]. In contrast to the Raman intensity and extracted continuum background, the dark-field scattering maxima showed no systematic dependence on molecular LUMO energy, indicating that the observed enhancement trend is not governed by plasmon resonance shifts[46] (Fig. 2a and Fig. S4). Together, the substituent-dependent Raman intensity and continuum background, combined with the absence of a corresponding dark-field trend, suggest the involvement of a chemical enhancement mechanism.

Within a simplified energetic framework, substrate-to-molecule charge transfer becomes favourable when the incident photon energy bridges the energy separation between the metal Fermi level and the molecular acceptor state. Considering the Au Fermi level near -5.3 eV[47,48], excitation with a 632 nm laser (1.96 eV) would suggest optimal resonance for molecular states located near -3.34 eV. Experimentally, however, the maximum enhancement is consistently observed at slightly lower energies around -3.5 eV. This offset can be rationalized by considering that CT-mediated Raman enhancement is not governed only by the incident photon energy but also involves outgoing and vibronically coupled resonance pathways. For the 1585 $cm^{-1}$ Stokes mode, the scattered photon energy is reduced by ~0.20 eV, giving an effective outgoing resonance condition near -3.54 eV, close to the observed optimum. Thus, the resonance maximum likely reflects vibronically assisted CT rather than a rigid Fermi-to-LUMO threshold. Although the –CN derivative lies close to this estimated resonance condition, its lower intensity relative to –CHO indicates that energetic alignment is necessary but not sufficient. The CT enhancement also depends on interfacial energetic renormalization[49] arising from several coupled effects, including metal–molecule hybridization[50], image-charge stabilization[49], work-function modification[51] within the confined plasmonic cavity, local electrostatic fields, and vibronic broadening of the charge-transfer transition[52]. Excitation slightly above the minimum energetic threshold is often favorable for efficient charge-transfer coupling in CT-mediated SERS systems[53]. Collectively, these effects can shift and broaden the effective CT resonance window relative to the simple one-electron approximation.

The behaviour of the –$NO_2$ derivative provides additional insight into the complexity of the enhancement mechanism. Although the molecule lies slightly beyond the apparent optimum resonance window, it retains comparatively strong intensity for the 1585 $cm^{-1}$ aromatic stretching mode. This anomalous behaviour is likely associated with enhanced electronic delocalization induced by the strongly conjugating nitro substituent, which increases the polarizability and vibronic coupling of the aromatic framework[54]. Interestingly, this anomaly is significantly less pronounced for the 1280 $cm^{-1}$ inter-ring stretching mode and the 1080 $cm^{-1}$ C–S vibration, indicating that the enhanced response of the 1585 $cm^{-1}$ mode originates primarily from substituent-induced modulation of the aromatic π-electron cloud rather than from a fundamentally different CT mechanism.

To evaluate whether the observed enhancement profile originates specifically from excitation-energy-dependent CT resonance, the excitation wavelength was subsequently shifted from 632 nm to 785 nm (Fig. S5). Upon this lower-energy excitation, the resonance-like dependence observed at 632 nm collapses almost entirely into a non-systematic intensity distribution across the molecular library (Fig. 2b). Using the Au Fermi level of approximately -5.3 eV, the 785 nm laser can access CT resonances near -3.72 eV based on the incoming photon energy. For Stokes Raman scattering, however, the effective outgoing resonance is further lowered by the vibrational quantum by approximately 0.20 eV, shifting the accessible resonance window to ~-3.92 eV. Because the calculated LUMO energies of the BPT derivatives lie between -3.60 and -3.16 eV, the molecular acceptor states remain outside or only at the edge of the accessible CT resonance window upon 785 nm excitation. Consequently, direct substrate-to-molecule charge injection becomes energetically unfavorable, suppressing the CT-assisted enhancement pathway and leaving a largely non-resonant SERS response. Interestingly, the $–NO_2$ derivative retains comparatively higher intensity even under 785 nm excitation. This behavior is likely associated with its proximity to the energetic threshold for weak near-resonant CT contributions[55]. Additionally, the highly delocalized electronic structure of the nitro-substituted biphenyl system may facilitate broader vibronic coupling and partial resonance broadening, allowing residual enhancement even under marginally off-resonant excitation conditions.

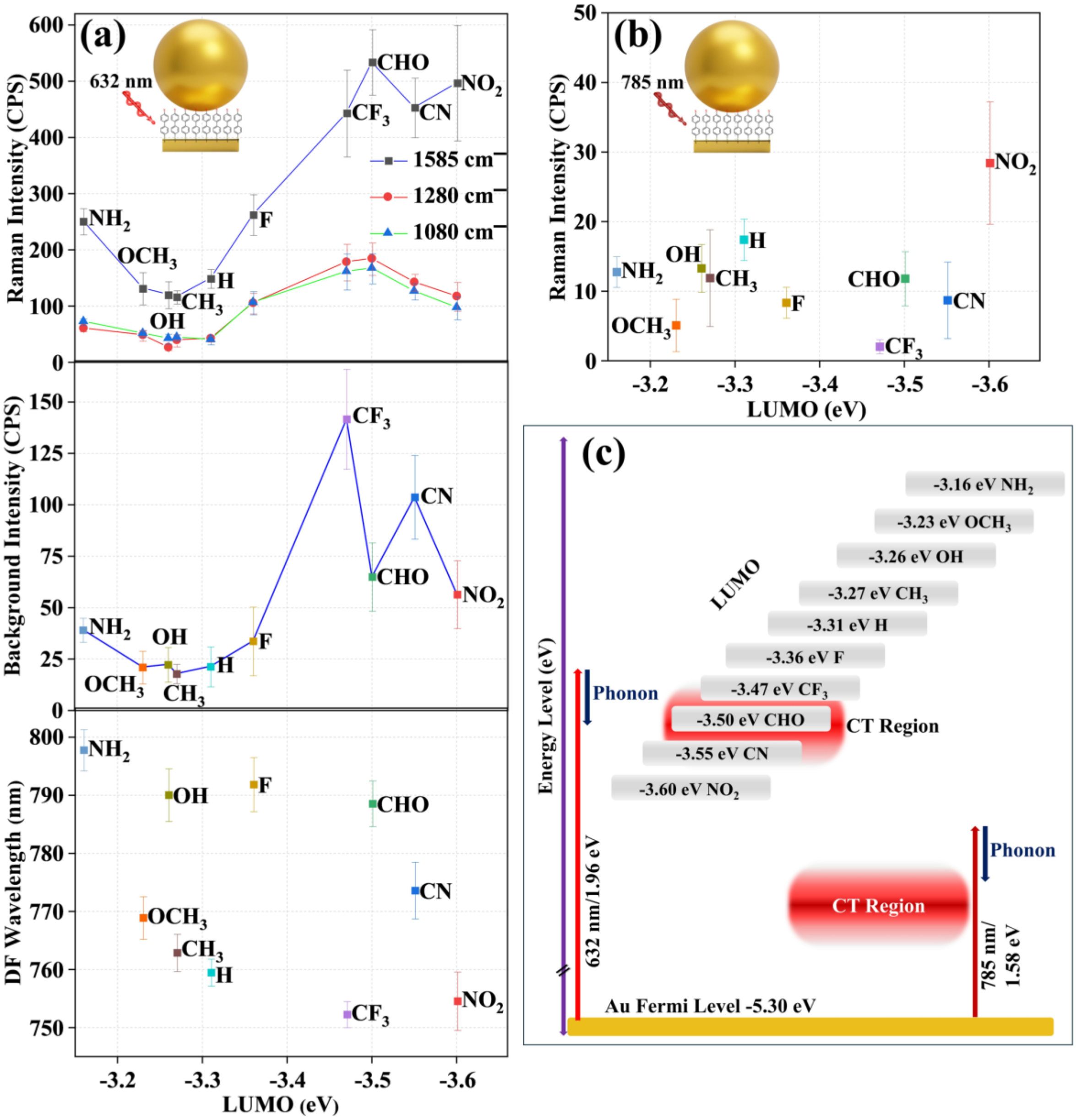

**Fig. 2 (a)** Correlation of calculated molecular LUMO energies with SERS intensities from Au NPoM junctions under 632 nm excitation. The upper panel shows the intensities of the 1585, 1280, and 1080 $cm^{-1}$ vibrational modes; the middle panel shows the extracted broad SERS background intensity; and the lower panel shows the corresponding dark-field scattering maxima. **(b)** Correlation between molecular LUMO energies and the 1585 $cm^{-1}$ SERS intensities under 785 nm excitation **(c)** Energy-level schematic illustrating substrate-to-molecule charge-transfer alignment in Au NPoM cavities. Under 632 nm excitation, the Au Fermi level can access the molecular acceptor states through an outgoing vibronic resonance window, whereas 785 nm excitation is energetically insufficient to produce efficient charge-transfer enhancement.

The overall behavior is summarized schematically in Fig. 2c. Under 632 nm excitation, the incident photon energy efficiently accesses a narrow subset of molecular acceptor states lying near the optimal interfacial resonance condition, producing strong CT-assisted enhancement. In contrast, 785 nm excitation lacks sufficient energy to effectively populate these states, thereby suppressing the CT channel and collapsing the resonance behavior. Together, these observations establish that Raman enhancement within the Au NPoM cavities is highly sensitive to frontier orbital alignment, with a narrowly defined energetic resonance condition between the metal Fermi level and the molecular LUMO.

### 2.3 Substrate Contributions and the Role of the Metal Fermi Level

Although the excitation-wavelength dependence strongly supports the presence of a resonant charge-transfer (CT) process, it does not by itself discriminate the spatial origin of the injected carriers within the NPoM junction, which could come from either the planar metallic substrate or the coupled nanoparticle. A third possibility is that the observed CT resonance is a collective effect involving the entire nanocavity construct. To decouple these contributions, a hybrid plasmonic architecture was constructed consisting of an Au planar substrate coupled to Ag nanoparticles, and SERS spectra were recorded. Under 632 nm excitation, the Au-substrate/Ag-nanoparticle hybrid system reproduces the similar resonant enhancement profile observed in the Au NPoM geometry (Fig. 3a and Fig. S6). Relative to the $–NH_2$ derivative, the –CHO molecule displays an enhancement of approximately five-fold for the 1585 $cm^{-1}$ vibrational mode, while also maintaining substantially stronger intensity than the electron-withdrawing $–NO_2$ derivative. Similar enhancement trends are consistently observed for the 1280 $cm^{-1}$ inter-phenyl stretching mode and the 1080 $cm^{-1}$ Au–thiolate-associated vibration. The persistence of the resonance maximum near the same LUMO energy despite substitution of the nanoparticle material provides strong mechanistic evidence that the dominant CT process originates from the metal to which the thiol group is bound, while no significant chemical enhancement occurs on the other side of the molecule.

Unlike Au, Ag exhibits a significantly shallower Fermi level around -4.35 eV[56,57]. Upon 632 nm excitation, photoexcited carriers generated within the Ag substrate can therefore access states near -2.54 eV. However, the calculated LUMO energies of the Ag-bound BPT derivatives remain substantially deeper, spanning approximately -2.96 to -3.36 eV. Consequently, the plasmonically generated carriers are energetically misaligned with the molecular acceptor states, preventing efficient resonant population of the LUMO manifold. To confirm this hypothesis, the entire plasmonic cavity was converted into a pure Ag NPoM system, and SERS spectra were recorded (Fig. S7). In contrast to the Au-based architectures, Ag NPoM cavities maintain high absolute Raman intensities, owing to the intrinsically strong electromagnetic enhancement provided by Ag plasmonics[58,59], however the resonant behaviour observed in the Au systems collapses into an irregular intensity distribution (Fig. 3b). This absence of correlation persists across multiple vibrational modes. Similarly, Ag NPoM measurements performed under 785 nm excitation also display a nearly flat and non-systematic intensity distribution (Fig. S8 and Fig. S9), further indicating the absence of a well-defined CT resonance process. Rather than producing a resonantly enhanced CT transition in Ag NPoM, the excitation effectively overshoots the energetically accessible molecular states, thereby suppressing selective substrate-to-molecule CT resonance (Fig. 3c). Taken together, these results demonstrate that the experimentally observed resonance behavior is governed predominantly by the electronic structure of the metal to which the molecule is chemically bound (here, through the thiol group) rather than by the post-deposited nanoparticle composition. The shallower Ag Fermi level suppresses the CT-assisted enhancement pathway irrespective of excitation wavelength.

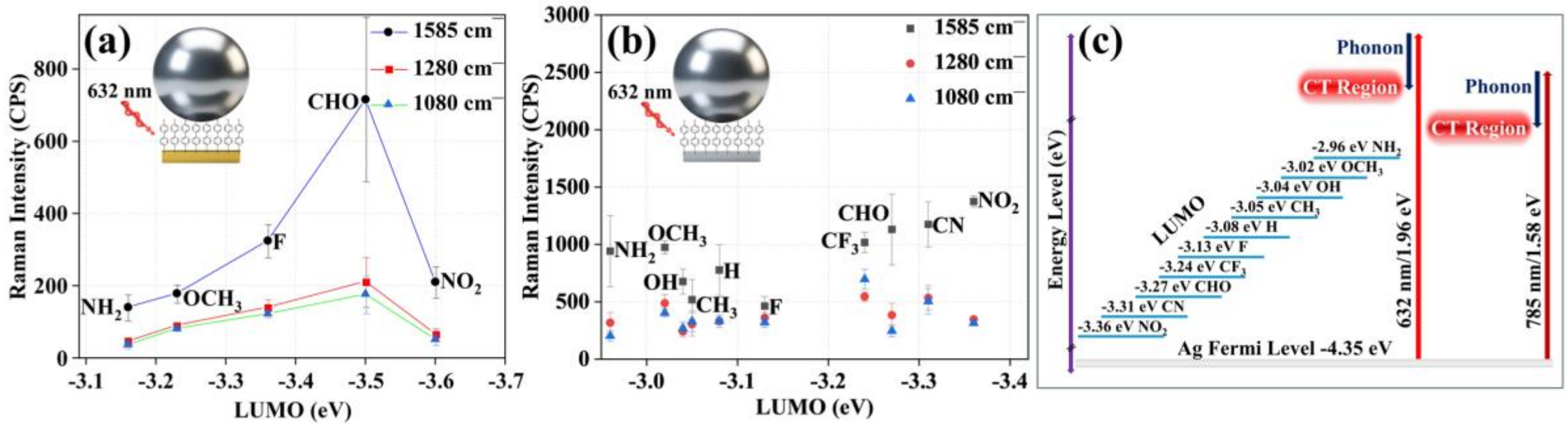

**Fig. 3 (a)** Correlation of calculated molecular LUMO energies with SERS intensities from Au-substrate/Ag-nanoparticle hybrid NPoM junctions under 632 nm excitation for the 1585, 1280, and 1080 $cm^{-1}$ vibrational modes. **(b)** Corresponding LUMO-dependent SERS intensities from Ag NPoM junctions under 632 nm excitation. **(c)** Energy-level schematic illustrating the mismatch between the Ag Fermi level and molecular LUMO states under both 632 and 785 nm excitation.

### 2.4 Validation via Predictive Molecular Design

Having established a clear empirical relationship between molecular orbital alignment and SERS enhancement, we next sought to determine whether the experimentally identified CT resonance condition could be used predictively to engineer molecular systems with enhanced optical response. Three additional molecular derivatives (PH, Ph_Sulfa, and N_Py) possessing structurally different backbones were designed and synthesized to target specific LUMO energies relative to the experimentally identified resonance window. DFT calculations predicted LUMO energies of approximately -3.37 eV for PH, -3.51 eV for Ph_Sulfa, and -3.70 eV for N_Py. The objective was to intentionally match the molecular acceptor state to the empirically determined resonance condition near -3.5 eV. Upon 632 nm excitation within the Au NPoM architecture, the experimentally measured Raman intensities closely follow the predicted trend (Fig. 4a and Fig. 4b). The Ph_Sulfa derivative exhibits much stronger enhancement than either PH or N_Py. Relative to PH, Ph_Sulfa demonstrates a higher that five-fold enhancement for the 1585 $cm^{-1}$ aromatic stretching mode. Similarly, Ph_Sulfa exhibits approximately 2.5-fold stronger intensity than the deeper N_Py derivative. The same energetic ordering is consistently preserved across the 1280 $cm^{-1}$ inter-phenyl stretching vibration and the 1080 $cm^{-1}$ Au–thiolate associated mode, indicating that the enhancement behavior is not restricted to a single vibrational transition.

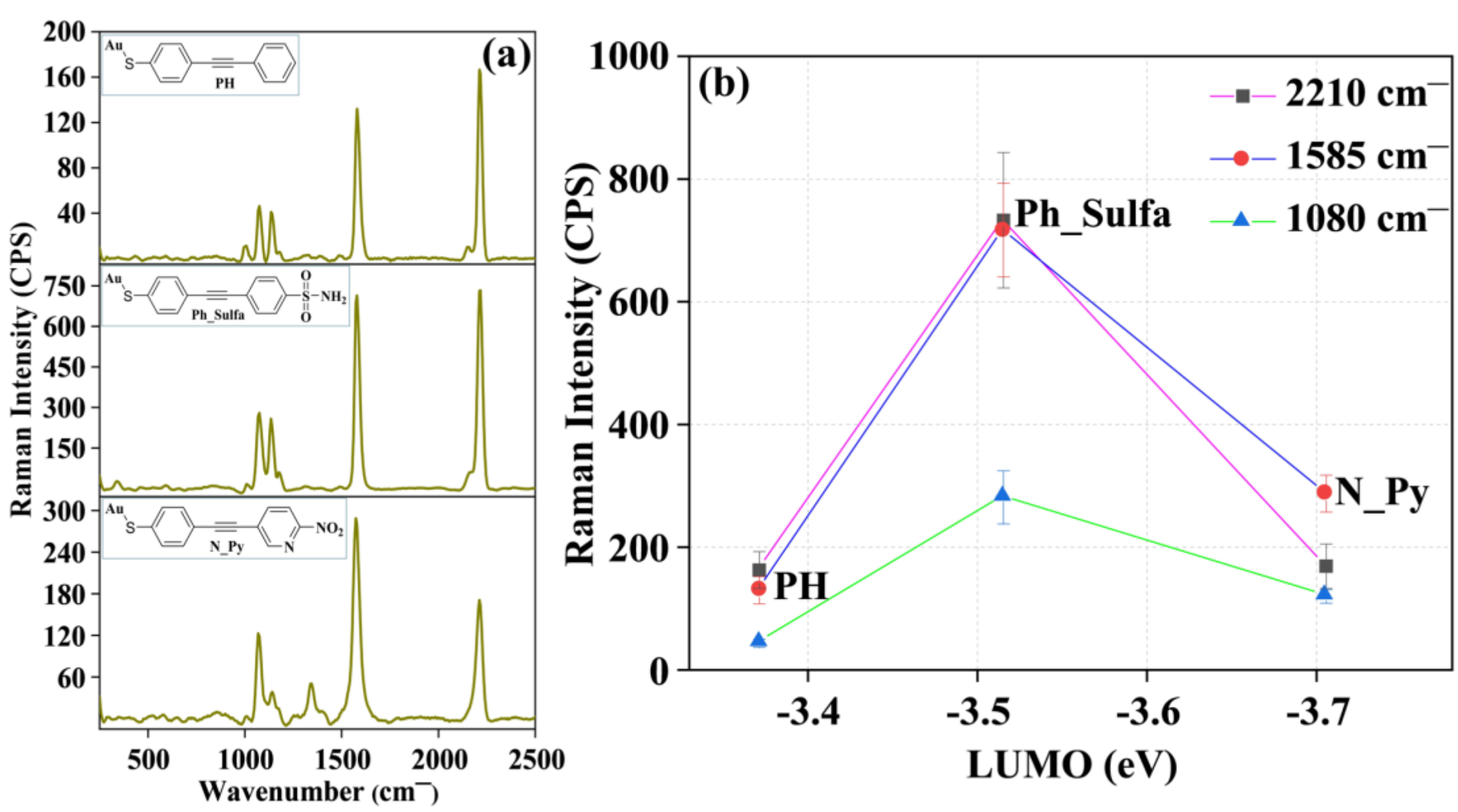

**Fig. 4 (a)** SERS spectra of PH, Ph_Sulfa, and N_Py assembled in Au NPoM junctions under 632 nm excitation. (b) Correlation of calculated molecular LUMO energies with SERS intensities for the 2210, 1585, and 1080 $cm^{-1}$ vibrational modes.

The predictive success of the Ph_Sulfa derivative demonstrates that the experimentally observed resonance condition is not unique to the original BPT framework or governed solely by substituent-specific effects. Instead, the enhancement maximum remains strongly correlated with energetic alignment relative to the substrate Fermi level, even when the molecular backbone itself is modified. This observation provides compelling evidence that frontier orbital alignment acts as the primary governing parameter controlling CT-assisted enhancement within the plasmonic cavity.

### 2.5 Sum Frequency Generation (SFG) and Thermodynamic Alignment

Having established design principles for achieving CT resonant enhancement in spontaneous, incoherent Raman scattering, we generalize the mechanism to coherent nonlinear spectroscopy. Specifically, we perform vibrational sum frequency generation (vSFG) spectroscopy[60], which is an intrinsically interface-selective technique probing regions lacking centrosymmetry[61]. It was observed that even in the absence of a plasmonic nanocavity, vSFG provides direct access to molecular layers and molecule-metal interfaces in more general contexts[62]. Here, we use the dual-resonant (at mid-infrared and visible wavelengths) nanoparticle-on-slit geometry[63,64], built around the same molecular monolayers as before, to acquire broadband vSFG spectra under tuneable continuous wave excitation from a quantum cascade laser.

The vSFG spectra acquired from the representative molecular subset show a pronounced enhancement of the aromatic C=C stretching mode near 1585 $cm^{-1}$ for the –CHO derivative (Fig. 5a, and Fig. S10). Quantitative comparison of the vSFG intensities shows that –CHO dominates the response, giving approximately four-fold higher intensity than $–NO_2$ and more than five-fold enhancement relative to $–NH_2$ (Fig. 5b). The recovery of the same molecular optimum observed in SERS supports a common interfacial CT resonance window near the –CHO LUMO. Importantly, vSFG probes this resonance through a different optical pathway than linear SERS. In Stokes SERS, enhancement can involve both incoming and outgoing resonance conditions but since the scattered photon is reduced in energy by the vibrational quantum, this makes the apparent CT sensitive to outgoing resonance effects. In vSFG, however, the relevant resonance is an incoming sum-frequency condition, i.e the visible photon and the resonant IR vibrational photon. Therefore, the vibrational energy is added to the visible excitation rather than subtracted (Fig. 5c). Using the Au Fermi level of −5.30 eV, the 785 nm visible excitation energy of 1.58 eV, and the vibrational energy of the 1585 $cm^{-1}$ mode of ~0.20 eV, the vSFG resonance condition is estimated as:

$$-5.30 \text{ eV} + 1.58 \text{ eV} + 0.20 \text{ eV} \approx -3.52 \text{ eV}$$

This value closely matches the calculated LUMO energy of the –CHO derivative, providing strong evidence that the dominant vSFG response arises from a vibronically assisted substrate-to-molecule CT resonance. The agreement is particularly significant because vSFG selectively probes only the highly ordered molecules directly participating in the interfacial nonlinear optical process[65]. In contrast, SERS samples the broader plasmonic hotspot ensemble, where local field heterogeneity, dipolar screening, cavity broadening, and structural disorder can broaden the apparent resonance window[66,67]. Together, the SERS and vSFG results indicate that the true interfacial CT resonance lies near -3.5 eV, with SFG revealing this alignment through an incoming sum-frequency resonance and SERS reflecting the related outgoing Raman resonance.

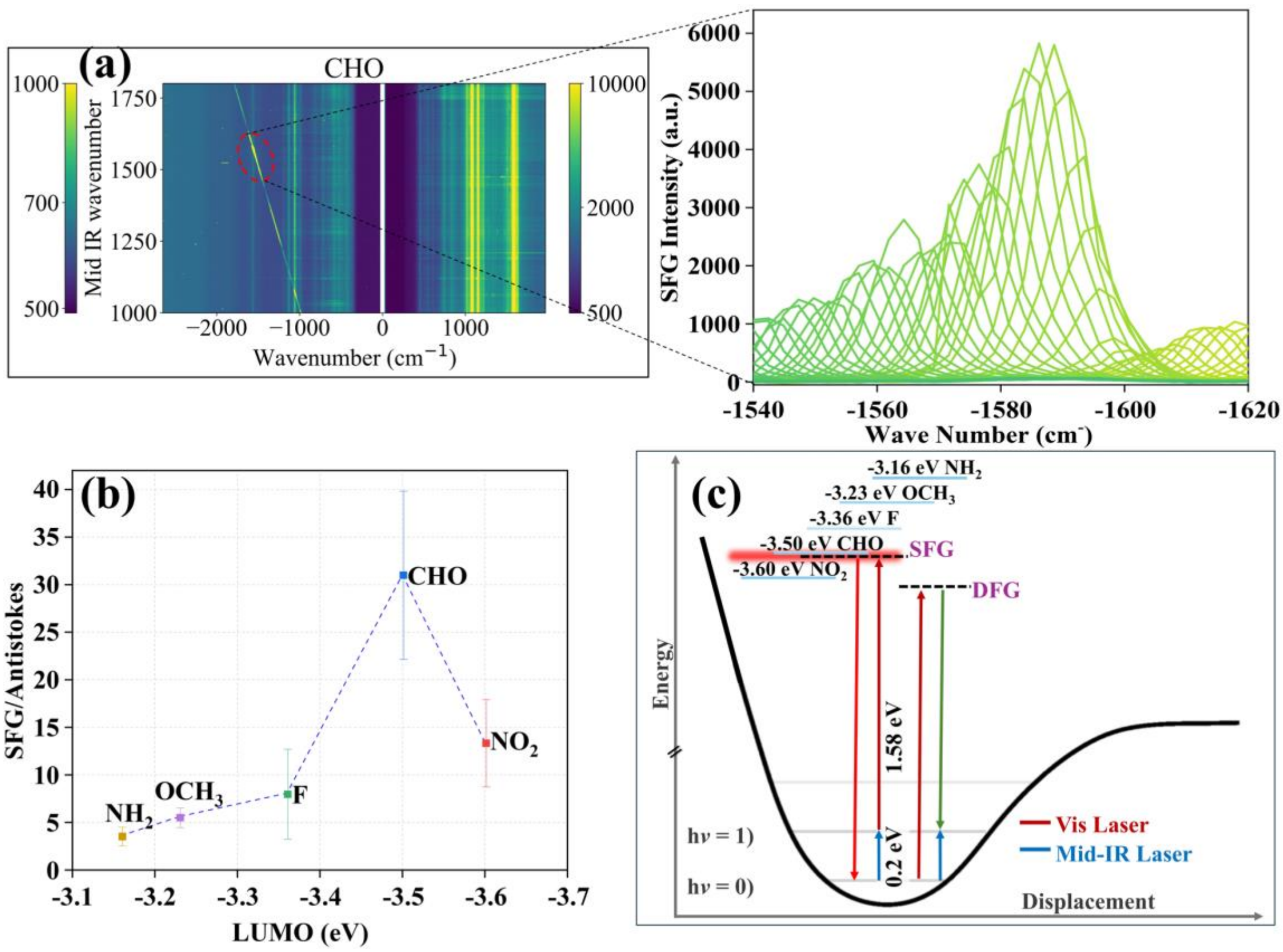

**Fig. 5 (a)** Mid-IR-scanned emission maps from Au NPoS cavities containing the representative –CHO spacer, acquired by tuning the IR excitation from 1800 to 1000 $cm^{-1}$ under fixed 785 nm visible excitation together with selected vSFG spectral region centered near the 1585 $cm^{-1}$ aromatic C=C stretching mode. **(b)** Ratio of the anti-Stokes Raman intensity at 1080 $cm^{-1}$ to the vSFG intensity at 1585 $cm^{-1}$ for the selected molecular subset, showing that the –CHO derivative exhibits the strongest relative nonlinear interfacial response. **(c)** Energy-level schematic for vSFG-mediated charge-transfer resonance, in which the combined energy of the 785 nm visible photon and the 6.3 μm IR photon (~0.20 eV, 1585 $cm^{-1}$) aligns the Au Fermi level with the molecular LUMO near the experimentally observed resonance window.

## 3. Conclusion

In this work, we systematically investigated the conditions governing interfacial charge-transfer enhancement within plasmonic nanoparticle-on-mirror nanocavities. Using a structurally consistent library of electronically different biphenylthiol derivatives, we experimentally mapped the relationship between molecular LUMO alignment and plasmon-assisted vibrational enhancement while minimizing competing structural effects within the molecular junction. Under 632 nm excitation in Au NPoM cavities, Raman enhancement exhibited a pronounced resonance-like maximum near a LUMO energy of approximately -3.5 eV, where the –CHO derivative displayed enhancement factors of ~3.5-fold for the 1585 $cm^{-1}$ mode, ~4.3-fold for the 1280 $cm^{-1}$ mode, and ~4.1-fold for the 1080 $cm^{-1}$ mode relative to the unsubstituted BPT derivative.

Excitation-energy-dependent measurements demonstrated that this energetic trend collapses under 785 nm excitation, confirming that efficient enhancement requires favorable energetic overlap between plasmonically accessible electronic states and the molecular LUMO manifold. Hybrid Au-substrate/Ag-nanoparticle architectures preserved the same resonance behavior as that of Au NPoM, whereas pure Ag NPoM systems exhibited no systematic dependence on molecular orbital energy, establishing the dominant role of the substrate Fermi level in governing interfacial charge injection. Rationally designed molecular derivatives further validated the predictive nature of the experimentally identified resonance condition, where the Ph_Sulfa derivative possessing a LUMO near -3.51 eV exhibited over five-fold enhancement compared to off-resonant analogues.

Furthermore, the successful prediction of maximal enhancement through rational orbital engineering establishes an important conceptual advance for plasmonic molecular design. Rather than treating chemical enhancement as an empirical or molecule-specific phenomenon, these results demonstrate that CT-assisted enhancement within NPoM cavities can be intentionally tuned through precise control of frontier orbital energetics. The Ph_Sulfa derivative therefore serves as a proof-of-concept for deterministic molecular engineering of plasmonic charge-transfer resonance, providing a potential framework for the rational development of optimized molecular junctions for ultrasensitive spectroscopy, plasmon-driven photochemistry, and charge-carrier-mediated interfacial processes.

Complementary SFG measurements revealed maximal interfacial enhancement under a doubly resonant condition corresponding to approximately –3.52 eV, in excellent agreement with the experimentally observed Raman. Collectively, these findings establish a quantitative framework for engineering charge-transfer resonance in plasmonic molecular junctions[68] and provide mechanistic insight relevant to plasmon-enhanced spectroscopy[69], charge-carrier photophysics[70,71], and nanoscale interfacial energy transfer[72,73].

## 4. Experimental

### 4.1 Synthesis

#### 4.1.1 Synthesis of Para-Substituted Biphenylthiol Derivatives (Scheme 1)

**4-(4,4,5,5-Tetramethyl-1,3,2-dioxaborolan-2-yl)benzenethiol (1)**

To a solution of (4-mercaptophenyl)boronic acid (20 mmol) in diethyl ether (60 mL) was added pinacol (30 mmol) and anhydrous sodium sulfate (60 mmol). The reaction mixture was stirred overnight at room temperature. The mixture was filtered, and the residue was washed with ethyl acetate (100 mL). The combined organic layers were washed with water and concentrated under reduced pressure to afford 4-(4,4,5,5-tetramethyl-1,3,2-dioxaborolan-2-yl)benzenethiol as a white solid in quantitative yield.

$^1$H NMR (400 MHz, DMSO-d6) δ 7.52 (dt, J = 8.0, 1.6 Hz, 2H), 7.29 (dt, J = 8.0, 1.6 Hz, 2H), 5.62 (s, 1H), 1.27 (s, 12H).

$^{13}$C NMR (101 MHz, DMSO-d6) δ 137.14, 135.00, 127.37, 83.56, 24.65.

**3-((4-(4,4,5,5-Tetramethyl-1,3,2-dioxaborolan-2-yl)phenyl)thio)propanenitrile (2)**

4-(4,4,5,5-Tetramethyl-1,3,2-dioxaborolan-2-yl)benzenethiol (5 mmol) was dissolved in THF (5 mL), followed by addition of triethylamine (5 mol%). The reaction mixture was cooled to 0 °C, and acrylonitrile (5.5 mmol) was added dropwise. The mixture was allowed to warm to room temperature and stirred overnight. Saturated aqueous $NH_4Cl$ was added, and the crude product was extracted with ethyl acetate. The organic phase was concentrated under reduced pressure and purified by column chromatography (heptane/ethyl acetate) to afford 3-((4-(4,4,5,5-tetramethyl-1,3,2-dioxaborolan-2-yl)phenyl)thio)propanenitrile as a white solid (76%).

$^1$H NMR (400 MHz, $CDCl_3$) δ 7.76 (d, J = 8.2 Hz, 2H), 7.36 (d, J = 8.2 Hz, 2H), 3.17 (t, J = 7.4 Hz, 2H), 2.61 (t, J = 7.4 Hz, 2H), 1.34 (s, 12H).

$^{13}$C NMR (101 MHz, $CDCl_3$) δ 137.21, 135.68, 129.21, 117.96, 84.07, 29.32, 24.95, 18.20.

#### 4.1.2 General Suzuki Cross-Coupling Procedure

The para-substituted biphenylthiol derivatives were synthesized via Suzuki–Miyaura cross-coupling with slight modification to a reported procedure[74]. In a dried microwave vial were added aryl halide (2 mmol), boronic ester intermediate (2 mmol), and $Pd(PPh_3)_4$ (5 mol%). The vial was evacuated and backfilled with nitrogen three times. Degassed dimethoxyethane (10 mL) and degassed aqueous $K_2CO_3$ solution (2 M, 5 mL) were added, and the sealed vial was stirred overnight at 90 °C. After cooling to room temperature, saturated aqueous $NH_4Cl$ was added and the reaction mixture was extracted with dichloromethane. The combined organic layers were concentrated under reduced pressure and purified by column chromatography (heptane/DCM) to afford the desired para-substituted biphenylthiol derivatives. Detailed characterization data, isolated yields, and spectral analyses for all derivatives are provided in the Supporting Information.

**Scheme 1:** Synthetic scheme of BPT derivatives

### 4.1.3 Synthesis of Alkyne-Containing Derivatives (Scheme 2)

**3-((4-Bromophenyl)thio)propanenitrile (3)**

4-Bromobenzenethiol (5 mmol) was dissolved in THF (5 mL), followed by addition of triethylamine (5 mol%). The reaction mixture was cooled to 0 °C, and acrylonitrile (5.5 mmol) was added dropwise. The reaction was allowed to warm to room temperature and stirred overnight. Saturated aqueous $NH_4Cl$ was added and the crude product was extracted with ethyl acetate. The organic layer was concentrated under reduced pressure and purified by column chromatography (heptane/ethyl acetate) to afford 3-((4-bromophenyl)thio)propanenitrile as a white solid (87%).

$^1$H NMR (400 MHz, $CDCl_3$) δ 7.46 (d, J = 8.1 Hz, 2H), 7.28 (d, J = 8.1 Hz, 2H), 3.11 (t, J = 7.2 Hz, 2H), 2.59 (t, J = 7.2 Hz, 2H).

$^{13}$C NMR (101 MHz, $CDCl_3$) δ 133.05, 132.64, 132.51, 122.09, 117.84, 30.43, 18.36.

**3-((4-((Triisopropylsilyl)ethynyl)phenyl)thio)propanenitrile (4)**

In a dried microwave vial were added 3-((4-bromophenyl)thio)propanenitrile (4 mmol), $Pd(PPh_3)_4$ (2 mol%), and CuI (5 mol%). The vial was evacuated and backfilled with nitrogen three times, followed by addition of degassed $Et_3N$ (10 mL) and triisopropylsilyl acetylene (6 mmol). The sealed vial was stirred overnight at 60 °C. After cooling, saturated aqueous $NH_4Cl$ was added, and the mixture was extracted with dichloromethane. The organic phase was concentrated under reduced pressure and purified by column chromatography (heptane/ethyl acetate) to afford 3-((4-((triisopropylsilyl)ethynyl)phenyl)thio)propanenitrile as an off-white solid (92%).

HRMS (ESI) m/z: [M + H]+ calcd for $C_{20}H_{30}NSSi$, 344.1869; found, 344.1872.

$^1H$ NMR (400 MHz, $CDCl_3$) δ 7.46–7.41 (m, 2H), 7.35–7.30 (m, 2H), 3.14 (t, J = 7.3 Hz, 2H), 2.60 (t, J = 7.3 Hz, 2H), 1.13 (s, 21H).

$^{13}C$ NMR (101 MHz, $CDCl_3$) δ 133.84, 132.97, 130.61, 122.98, 117.89, 106.22, 92.41, 30.03, 18.80, 18.30, 11.43.

**3-((4-Ethynylphenyl)thio)propanenitrile (5)**

To a solution of 3-((4-((triisopropylsilyl)ethynyl)phenyl)thio)propanenitrile (1 mmol) in DCM (3 mL) was added dropwise TBAF (1 M in THF, 1.5 mmol). The reaction mixture was stirred at room temperature for 1 h. The solvent was removed under reduced pressure and the crude product was purified by column chromatography (heptane/ethyl acetate) to afford 3-((4-ethynylphenyl)thio)propanenitrile as a brown oil that solidified upon standing to give a light-brown powder (49%).

$^1H$ NMR (400 MHz, CDCl3) δ 7.48–7.41 (m, 2H), 7.36–7.30 (m, 2H), 3.16 (t, J = 7.3 Hz, 2H), 3.13 (s, 1H), 2.61 (t, J = 7.3 Hz, 2H).

$^{13}C$ NMR (101 MHz, $CDCl_3$) δ 134.74, 133.03, 130.30, 121.33, 117.85, 82.90, 78.54, 29.73, 18.31.

#### 4.1.4 General Sonogashira Coupling Procedure

In a dried microwave vial were added aryl halide (1 mmol), 3-((4-ethynylphenyl)thio)propanenitrile (1.05 mmol), $Pd(PPh_3)_4$ (5 mol%), and CuI (10 mol%). The vial was evacuated and backfilled with nitrogen three times. Degassed THF (5 mL) and degassed $Et_3N$ (5 mL) were added, and the sealed reaction mixture was stirred overnight at 70 °C. After cooling, saturated aqueous $NH_4Cl$ was added and the mixture was extracted with dichloromethane. The combined organic layers were concentrated under reduced pressure and purified by column chromatography (heptane/DCM) to afford the desired alkyne-containing derivatives. Detailed characterization data, isolated yields, and spectral analyses are provided in the Supporting Information.

**Scheme 2:** Synthetic scheme of alkyne containing derivatives

### 4.2 Sample Preparation

The devices were fabricated on double-polished silicon wafers with a thickness of 380 μm. To promote robust adhesion between the gold and silicon interfaces, a 5 nm aluminum (Al) adhesion layer and a subsequent 150 nm gold (Au) thin film were deposited via thermal evaporation at a tightly controlled rate of 0.5 nm/s. Following deposition, the substrates were spin-coated with PMMA 950 A4 resist at 2000 rpm and soft-baked to achieve a uniform resist profile. Nanoslit patterns were then defined using electron-beam lithography (EBL).

After resist development, the exposed gold layer was structured using a two-step collimated ion-beam etching process. The primary etching step was conducted at a 10° incidence angle to generate V-shaped trenches featuring slanted sidewalls. This was immediately followed by a secondary etching step executed at a 10° grazing angle to suppress redeposition fencing, thereby ensuring sharp, well-defined slit edges.

To complete the fabrication, the substrates were immersed in Remover 1165 at 60 °C for 24 hours to strip the remaining PMMA resist. Finally, a 10-second piranha solution dip was performed to eliminate residual organic contaminants, followed by a thorough deionized water rinse. Scanning electron microscopy (SEM) confirmed highly uniform final nanoslit dimensions, yielding a consistent length of 1.9 μm and a width of 140 nm.

Template-stripped gold substrate were prepared cleaning a Silicon wafer with Piranha solution for 5 minutes and subsequent 1min 02 plasma surface activation. A 150 nm gold (Au) thin film was deposited via thermal evaporation at a tightly controlled rate of 0.5 nm/s. 1x1 cm glass slide are left over night in ethanol for properly cleaning, later thanks by N81 epoxy glue the glass slides are fixed to the gold layer and cured for 15 min. Finally, the samples are peeled just before the experiment with sticky tape, transferring the flatness of the silicon wafer to the gold surface.

### 4.3 Substrate Functionalization

To deprotect the biphenyl derivatives, molecules were dissolved in THF achieving a concentration of 10mM which were then diluted to 1mM, with addition of ethanolic KOH solution while ensuring that the final pH is kept around 10.

To prepare self-assembled monolayers (SAM) NPoM, the template-stripped gold film was first incubated in the 1mM thiol THF/ethanol solution overnight. The substrate is rinsed then with ethanol and THF, followed by quick sonication for 5s to remove molecules adsorbed on top of the monolayer. The sample is again rinsed with ethanol and dried with a nitrogen flow. Once dry, an aqueous solution of 80 nm spherical Au nanoparticles (Nanopartz) or Ag (abcr GmbH) was drop-cast onto the substrate. After 5 min, the sample was rinsed with deionized water and dried using a nitrogen gun.

Nanoparticle functionalization was carried out by mixing 900 µL of citrate-coated Au nanoparticles (1 OD solution from Nanopartz) with 100 µL of 1 mM thiol solution. The mixture was sonicated for 1 h and then left to rest for an additional hour. It was subsequently centrifuged at 4800 RPM for 10 min to induce nanoparticle precipitation. The supernatant was discarded, and the precipitate was resuspended in deionized water. Redispersion was facilitated by sonication for 10 min. This washing procedure was repeated two additional times, using ethanol after removal of the supernatant, and finally once more using deionized water. To verify that no free thiols are present in the solution, after drop-casting the functionalized nanoparticles on the nanoslits, the remaining solution was centrifuged again, the supernatant solution was dried on a template-stripped Au film. Once dried, citrated-coated Au nanoparticles are drop-cast, and the SERS signal from the so formed NPOM is measured to confirm that no thiols are present

### 4.4 Optical Setup

The experimental setup consists of a confocal microscope designed to perform Raman and DF scattering spectroscopy[33]. The laser exposure is carried out by a continuous wave (CW) Titanium Sapphire (Ti:Sapph) laser from Spectra Physics (Model 3900, pumped by Millenia 5W) tuned at 785 nm. The laser line is filtered using a bandpass filter. Afterward, a lambda-half plate (LH), a polarizing beam splitter (PBS), and a sliding neutral density filter (ND) are used to control the power of the laser beam. The LH and the PBS allow for decreasing the laser power in large steps, while the ND enables finer tuning. A pellicle beam splitter (P) then reflects about 10% of the incident laser power toward a high NA objective (Olympus MPLFLN 100× 0.9NA) used to focus the laser on the NPoM and collect the Raman signal from the nanocavity-coupled molecules. HeNe laser excitation at 633 nm is delivered similarly.

The DF measurements are performed using a supercontinuum (NKT Photonics SuperK Compact) white light source focused on the sample by a secondary objective (Olympus Plan N 4× 0.1NA) at an angle of 81° from the normal to the sample plane. To use the supercontinuum as an interference-free white light source, its transverse spatial coherence length must be reduced. This is achieved by a rotating diffuser. For all the measurements, the signal from the sample is collected by the high NA objective. The acquired signal is split, and a small portion is used to image the sample plane while the rest is passed through spectral filters and finally recorded using a spectrograph (Andor Kymera 193i) with a CCD camera (Andor iDus).
Single-cavity NPoS O-PTIR and SFG spectra were measured via a commercial mIRage LS instrument (Photothermal Spectroscopy Corp) in co-propagation mode by scanning the mid-IR beam from 1000 to 1800 $cm^{-1}$.

**4.5 DFT Calculation**

According to the literature, modelling a single thiol molecule by replacing the thiol hydrogen with a gold atom in the simulation improves the DFT description of the experimental SERS spectra of thiol SAMs, leading to better agreement between theory and experiment[40]. Geometry optimization and vibrational frequency calculations were performed using density functional theory (DFT) with the B3LYP functional and Def2-TZVPP basis set in Gaussian on the DTU HPC server. Grimme's D3BJ dispersion correction was included. Geometry optimizations and Raman frequency calculations were carried out using an ultrafine integration grid without symmetry constraints. Optimized structures were confirmed as minima by the absence of imaginary frequencies.

**CRediT authorship contribution statement**

**Hasher Irshad:** Conceptualization, Methodology, Writing – original draft, Formal analysis. **Francesco Ciccarello:** Methodology, Formal analysis. **Konstantin Malchow:** Writing – review & editing, SFG Measurements, Formal analysis, Software. **Christophe Galland:** Validation, Supervision, Writing – review & editing. **Katrine Qvortrup:** Validation, Supervision, Funding, Writing – review & editing.

**Declaration of Competing Interest**

The authors declare that they have no known competing financial interests or personal relationships that could have appeared to influence the work reported in this paper.

**Acknowledgments**

The author (Katrine Qvortrup) is very grateful to DTU for providing financial support vide Grant No. 109005. This work received funding from the European Union's Horizon 2020 research and innovation program under Grant Agreement No. 820196 (ERC CoG QTONE) and from the Swiss National Science Foundation (grant number 214993). During the preparation of this work, the author(s) used ChatGPT to improve readability by enhancing sentence structure and correcting grammatical errors. After using this tool/service, the authors reviewed and edited the content as needed and take full responsibility for the content of the published article.

**Data Availability**

All the data needed is part of the manuscript or provided in supplementary information.